\pdfoutput=1

\documentclass[sigconf,10pt]{acmart}
\renewcommand\footnotetextcopyrightpermission[1]{} 
\usepackage{geometry}
\def\BibTeX{{\rm B\kern-.05em{\sc i\kern-.025em b}\kern-.08emT\kern-.1667em\lower.7ex\hbox{E}\kern-.125emX}}
    
\setcopyright{none}
\acmPrice{15.00}
\acmDOI{}
\acmISBN{}
\acmYear{}
\copyrightyear{}

\usepackage{todonotes}
\usepackage{booktabs}
\usepackage{balance}

\newif\ifcommenton
\commentontrue
\ifcommenton
\newcommand{\paras}[1]{\textcolor{blue}{PJ: #1}}
\newcommand{\simon}[1]{\textcolor{cyan}{XMO: #1}}
\newcommand{\ajay}[1]{\textcolor{red}{AJ: #1}}
\newcommand{\joey}[1]{\textcolor{olive}{JG: #1}}
\newcommand{\alexey}[1]{\textcolor{magenta}{AT: #1}}
\else
\newcommand{\alexey}[1]{}
\newcommand{\paras}[1]{}
\newcommand{\simon}[1]{}
\newcommand{\ajay}[1]{}
\newcommand{\joey}[1]{}
\fi

\begin{document}
\sloppy

%
\title{The OoO VLIW JIT Compiler for GPU Inference\vspace{-0.1in}}

%
\author{Paras Jain, Xiangxi Mo, Ajay Jain$^\ddagger$, 
Alexey Tumanov, Joseph E. Gonzalez, Ion Stoica}

\affiliation{
\vspace{-0.125in}
UC Berkeley, MIT$^\ddagger$}
\renewcommand{\shortauthors}{Jain, et al.}

%
\begin{abstract}
Current trends in Machine Learning~(ML) inference on hardware accelerated devices (e.g., GPUs, TPUs) point to alarmingly low utilization. As ML inference is increasingly time-bounded by tight latency SLOs, increasing data parallelism is not an option. The need for better efficiency motivates GPU multiplexing. Furthermore, existing GPU programming abstractions force programmers to micro-manage GPU resources in an early-binding, context-free fashion.
We propose a VLIW-inspired Out-of-Order (OoO) Just-in-Time (JIT) compiler that \textit{coalesces} and \textit{reorders} execution kernels at runtime for throughput-optimal device utilization while satisfying latency SLOs. We quantify the inefficiencies of space-only and time-only multiplexing alternatives and demonstrate an achievable 7.7x opportunity gap through spatial coalescing.



\end{abstract}

%
%

%

%

\maketitle

\begin{figure}[t]
    \centering
    \includegraphics[width=
    \columnwidth]{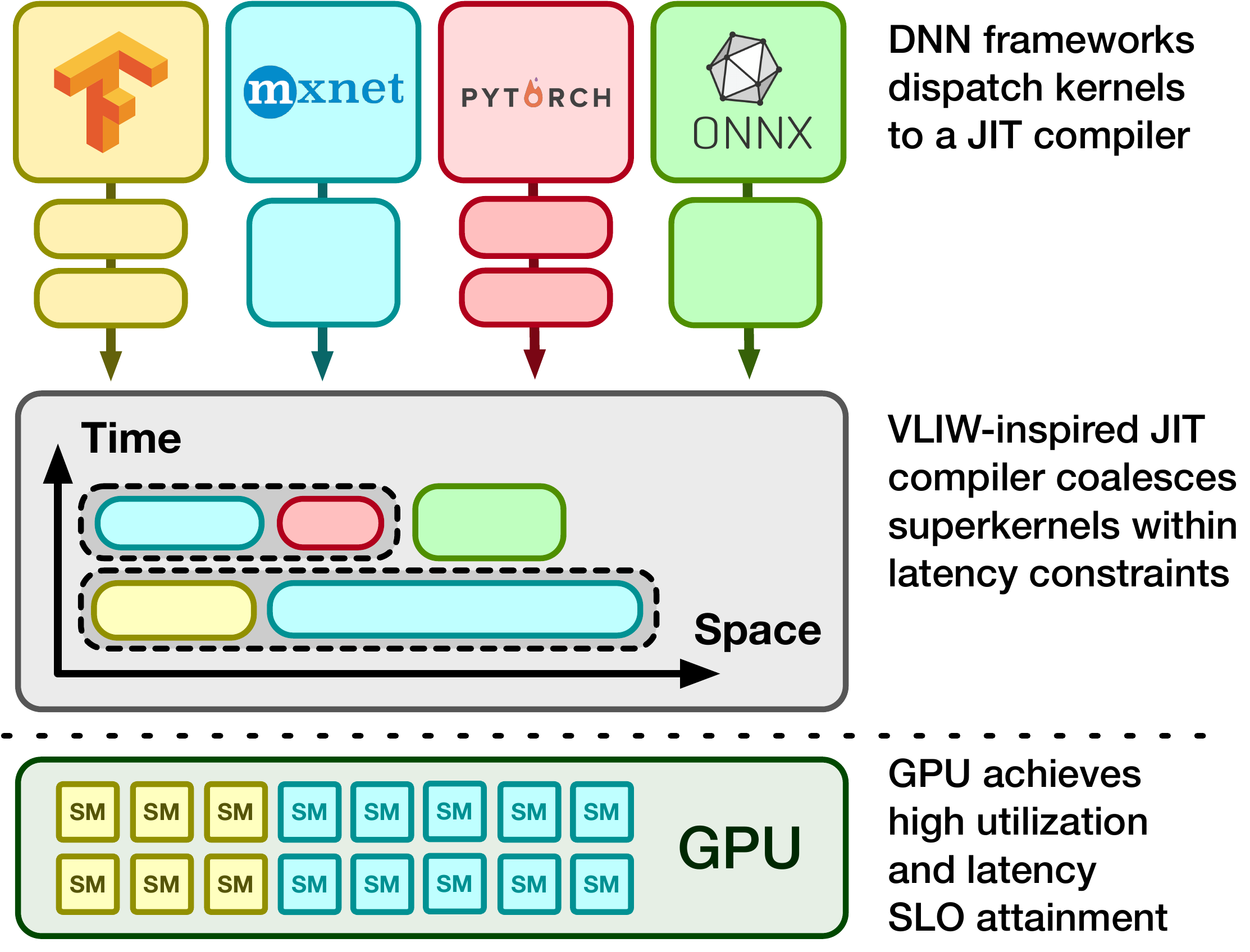}
    \caption{\small{
    A proposed OoO VLIW JIT compiler for on-GPU inference \textit{coalesces} and \textit{reorders} heterogeneous kernels from multiple streams of execution, effectively creating an efficient space-time schedule for on-GPU execution.
    Coalesced kernel will increase GPU compute and memory utilization. Interleaving multiple execution streams extracts OoO parallelism and reorders execution to fit in latency SLO budgets.
    }}
    \label{fig:teaser_figure}
    \label{fig:intro:frontpage}
\end{figure}

\section{Introduction}
\label{sec:intro}
As deep learning is deployed in applications ranging from video monitoring to language translation, there is an emerging need for parallel hardware accelerators to support inference.
Deep learning inference needs to scale to billions of queries per day and is rapidly outpacing training in datacenters~\cite{hazelwood2018applied}. 
Amazon estimates that 90\% of production ML infrastructure costs are for inference, not training~\cite{reinvent}.



While there are numerous specialized inference processors, the widespread availability of GPUs and their support for general deep learning models renders them indispensable for inference. By leveraging substantial parallelism, high memory bandwidth, and tensor acceleration, GPUs quickly evaluate deep neural networks.

\begin{figure}[t]
    \centering
    \includegraphics[width=\columnwidth]{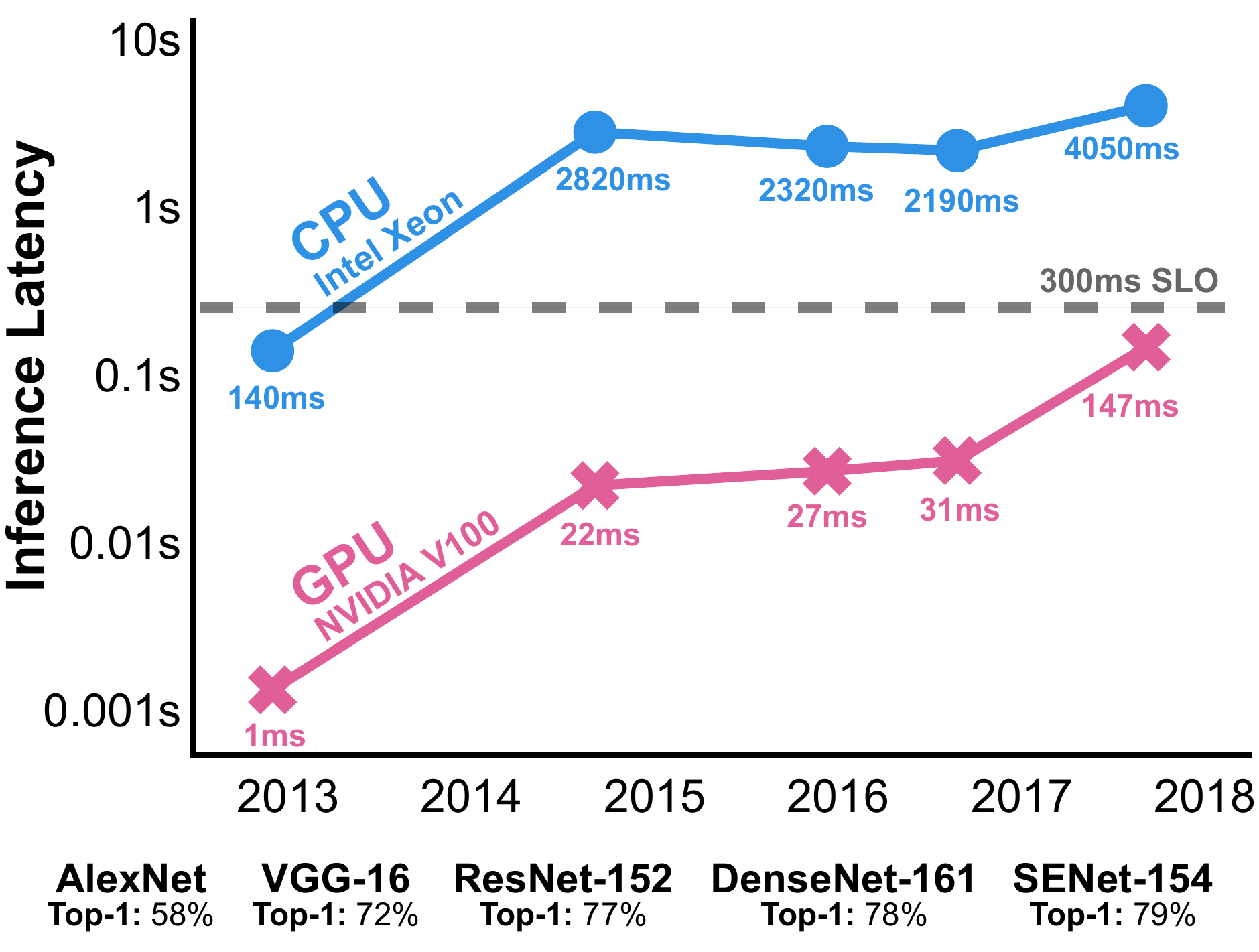}
    \caption{\small{
    DNN model complexity and inference latency is increasing over time on CPUs and GPUs. Most models fail to meet the 300ms latency SLO on a CPU. To meet low latencies, GPUs are seeing widespread use for DNN inference.
    }}
    \label{fig:intro:latency}
\end{figure}

Interactive inference workloads typically have tight latency objectives, especially when revenue-critical services issue inference queries.
We note a concerning trend that inference latency on a CPU has been on a rise~(\autoref{fig:intro:latency}); the state-of-the-art SENet-184~\cite{hu2018senet} model takes $4.1\textrm{s}$ for a single CPU inference. 
As models grow larger over time, CPUs will continue to struggle to serve interactive model serving workloads. With real revenue costs associated with user-facing latency~\cite{hamilton_2009, schurman_brutlag}, GPUs will remain a favorite for inference.

While some training workloads continue to scale and can often easily saturate modern GPUs, ML \textit{inference} has distinctly different performance requirements that often result in poor GPU utilization, given the current GPU programming abstractions.
In practice, online inference queries often cannot realize the high levels of parallelism that offline iterative minibatch training achieves, leading to poor GPU utilization. 
AWS reports \texttt{p3} GPU instances are only 10-30\% utilized~\cite{reinvent}. Low resource-efficiency is not isolated to GPUs as Google's Tensor Processing Unit reports a 28\% mean utilization~\cite{tpuISCA17}.

We propose an Out-of-Order (OoO) Just-in-Time (JIT) GPU kernel VLIW-inspired compiler for DNN inference~(\autoref{fig:intro:frontpage}). VLIW refers to computer architecture designed to extract instruction level parallelism (ILP) by packing multiple mutually-independent instructions that utilize different arithmetic processing units into a single large instruction word. VLIW design pushed the complexity of superscalar execution to the compiler while trying to keep the hardware simple. 
Analogously, we can improve accelerator utilization by reordering and packing kernels on-the-fly.  We borrow inspiration from VLIW by coalescing execution kernels into superkernels that can more fully leverage a large pool of massively parallel GPU compute units. Out-of-order execution between independent execution streams then increases efficiency while meeting latency deadlines.

Two reasons often attributed to the VLIW's failure are: (a) overwhelming multitude of operations to coalesce, making it hard for the compiler to solve the packing problem; (b) difficulty with extracting enough instruction level parallelism such that instructions don't have data dependencies between them. Despite its history, we believe the VLIW approach holds promise for GPU inference because 
(a) the set of operations to coalesce is restricted largely to algebraic tensor operations (e.g., matrix multiplies), 
(b) ability to leverage multiple inference streams, which, by construction, consist of mutually independent operations, and
(c) the shape of execution kernels is adjustable, while VLIW compilers operated on immutable instructions. 

We emphasize that the JIT compiler is \textit{dynamic}. It operates on multiple streams of execution (analogous to instruction streams) in real-time, coalescing and reordering the operations in GPU space-time~(\autoref{fig:intro:frontpage}). To the best of our knowledge, this is the first such proposition for on-GPU DNN inference.



\nocite{krizhevsky2012imagenet}\nocite{Simonyan14c}\nocite{he2016deep}\nocite{huang2017densely}\nocite{hu2018senet}

\section{DNN Inference Requirements}
\label{sec:reqs}

Deep learning workloads like object detection, speech recognition, and natural language processing are being rapidly deployed in datacenters. 
These models are computationally intensive demanding 10s of GFLOPS of computation per query and a single model can easily see millions of inference queries per day.
Facebook reports that its compute requirements for DNN inference have increased 3.5x in just 2 years~\cite{MaximNaumov_facebook_codesign}.

Typical applications see both batch inference queries and interactive latency-sensitive queries.
Batch inference queries have no strict latency deadline and are primarily concerned with maximizing throughput. System designers aim to optimize the \textit{cost-per-query} for batch inference.
However, interactive inference queries have strict latency deadlines. 
With a real revenue cost for increased tail latencies in user-facing applications~\cite{schurman_brutlag}, system designers aim to control the \textit{p99 latency} for online model serving. Typical latency SLOs can vary from 10ms for search ranking to several seconds for explicit content recognition~\cite{tpuISCA17, hazelwood2018applied, MaximNaumov_facebook_codesign}.
Furthermore, to meet interactive latency requirements, these models must run on specialized hardware accelerators (typically GPUs).

Interactive inference queries present a challenging dilemma for system designers. 
In order to meet their strict deadlines, small batch-sizes must be used~\autoref{fig:intro:batchsize}.
However, those small batch-sizes lead to poor resource-utilization and therefore high-costs under heavy load.

\section{An emerging utilization gap}
\label{sec:utilgap}

Deep learning has motivated the design of specialized hardware optimized for its predictable yet computationally expensive workloads. Recent accelerators are being deployed in the datacenter~\cite{tpuISCA17, dean2017hotchips, graphcore, Dally_efficient_methods_dl, groq, nvidia_dgx2, nvidia_t4} and the edge~\cite{jetson_xavier, edgetpu, appleneuralengine, hikey980, tesla_ai_chip}.

\begin{figure}[t]
    \centering
    \includegraphics[width=\columnwidth]{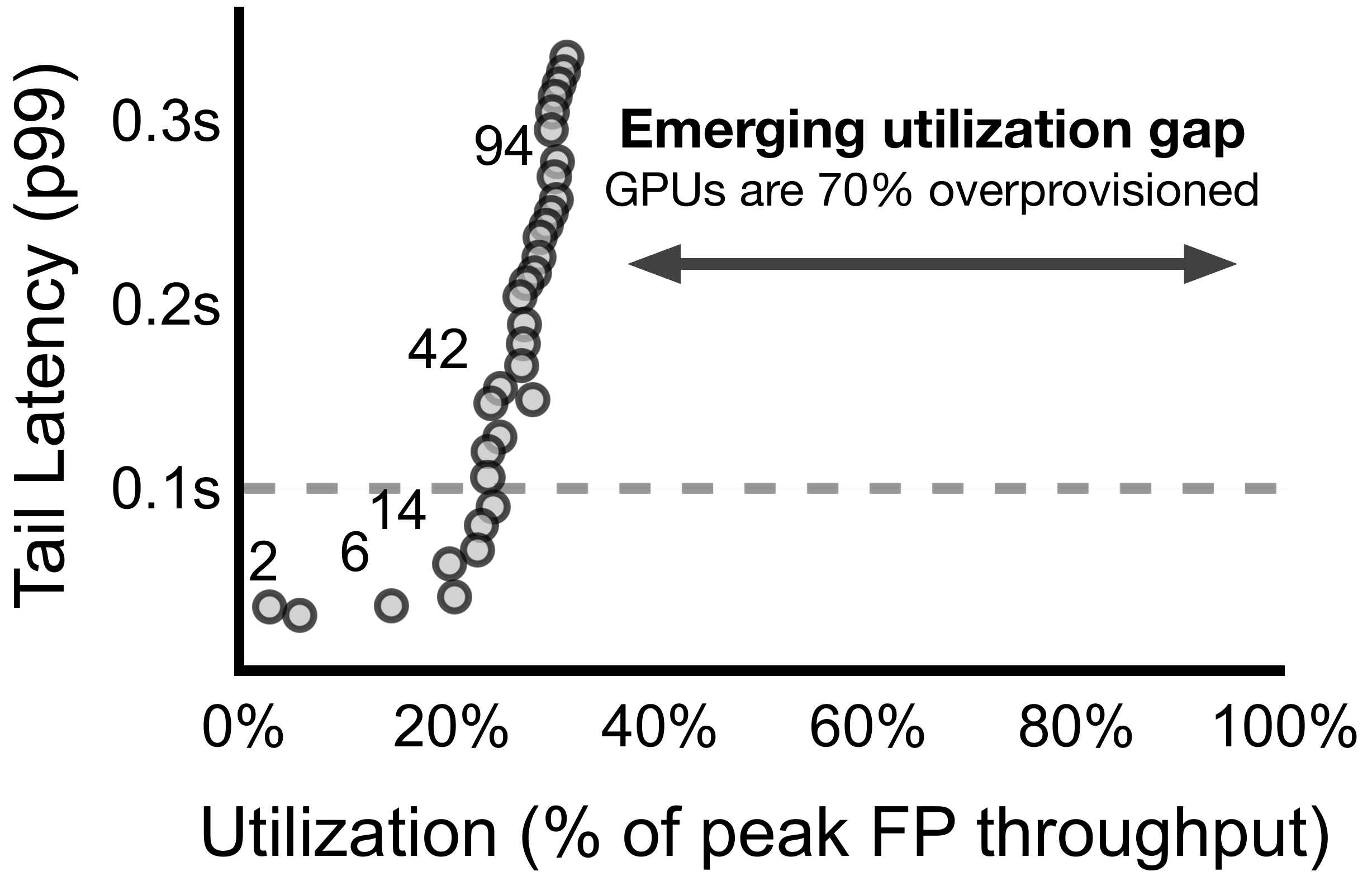}
    \caption{\small{
        An emerging utilization gap: In order to meet latency SLOs, small batch sizes must be used, resulting in low GPU utilization. 
        We profile ResNet50 across many batch-sizes (annotated above points) and see that single model inference will not fully utilize the GPU.
        }}
    \label{fig:intro:batchsize}
\end{figure}


However, throughput-optimized inference accelerators see low utilization when serving interactive DNN workloads. Throughput-oriented accelerators like GPUs and the TPU require substantial available parallelism to achieve peak throughput.

We observe an emerging \textit{utilization gap} between hardware peak throughputs and actual observed performance. In ~\autoref{fig:intro:batchsize}, we benchmark ResNet-50 on NVIDIA's V100 GPU. At interactive latencies, throughput is less than 25\% of peak. Larger batch sizes struggle to achieve 40\% of NVIDIA's advertised 15.7 TFLOPS throughput. DNN hardware accelerators suffer similar issues --- the TPU v1 sees just 8.2\% utilization for text processing workloads~\cite{tpuISCA17}.

Why are throughput-oriented accelerators underutilized during DNN inference?

First, \textit{tight latency objectives limit available parallelism} during inference. \autoref{fig:intro:batchsize} demonstrates that ResNet-50 inference on the NVIDIA V100 GPU achieves under 30\% utilization. 
Kernels with small batch sizes have a low arithmetic intensity and thus poor peak floating-point throughput by the roofline model~\cite{williams2009roofline}. 
Individually, these kernels do not expose sufficient parallelism to saturate all the cores on modern GPUs.

Second, \textit{resources must be provisioned for peak demand} rather than the average. As requests arrive stochastically, occasional bursts in request volume require over-provisioning accelerator resources. 
For an organization, peak load provisioning common in deep learning inference pipelines like those in~\cite{crankshaw-inferline} leads to excess purchases several times what average load would require.

  \begin{figure}[tb]
    \centering
    \includegraphics[width=\columnwidth]{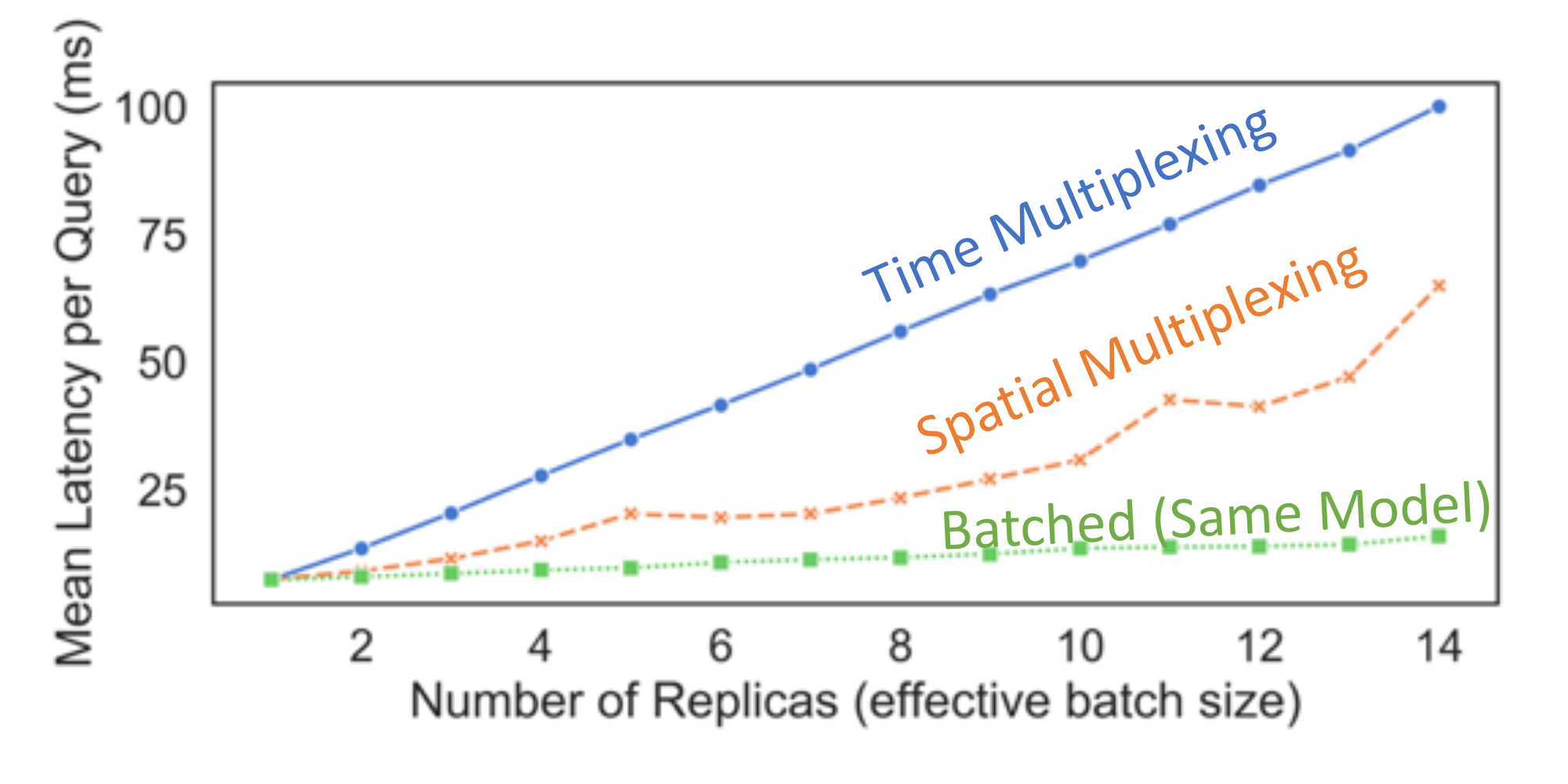}
    \caption{\small{
    Mean latency for 1 to 15 replicas of ResNet-50 on a V100 GPU. 
    Time multiplexing is not resource-efficient as it is dramatically slower than batched inference.
    Spatial multiplexing has unpredictable performance still-degraded from batched inference. 
    }}
    \label{fig:eval:inefficient}
  \end{figure}

Third, growth in \textit{compute throughput outpaces memory bandwidth}. Memory bandwidth no longer scales with increasing parallelism and compute throughputs~\cite{wulf1995hitting, asanovic2009view, rogers2009scaling}, and the ratio of compute throughput to memory bandwidth is rising rapidly. For GPUs, op to byte ratios have risen from 18 with the K80 to 139 for the V100.
Specialized hardware fares even worse; Google's TPUv2~\cite{dean2017hotchips} has an op to byte ratio of 300. We estimate AWS Inferentia~\cite{jameshamilton_inferentia} has op to byte ratio of almost 500.

These factors will lead to continued underutilization of DNN accelerators. Ultimately, we believe intelligent software-level scheduling will enable more efficient hardware accelerators. The VLIW-inspired approach we propose is one of many potential optimizations that operate at a higher level than hardware.
\section{Ineffective GPU multiplexing }
\label{sec:multiplex}

 \begin{figure}[tb]
    \centering
    \includegraphics[width=\columnwidth]{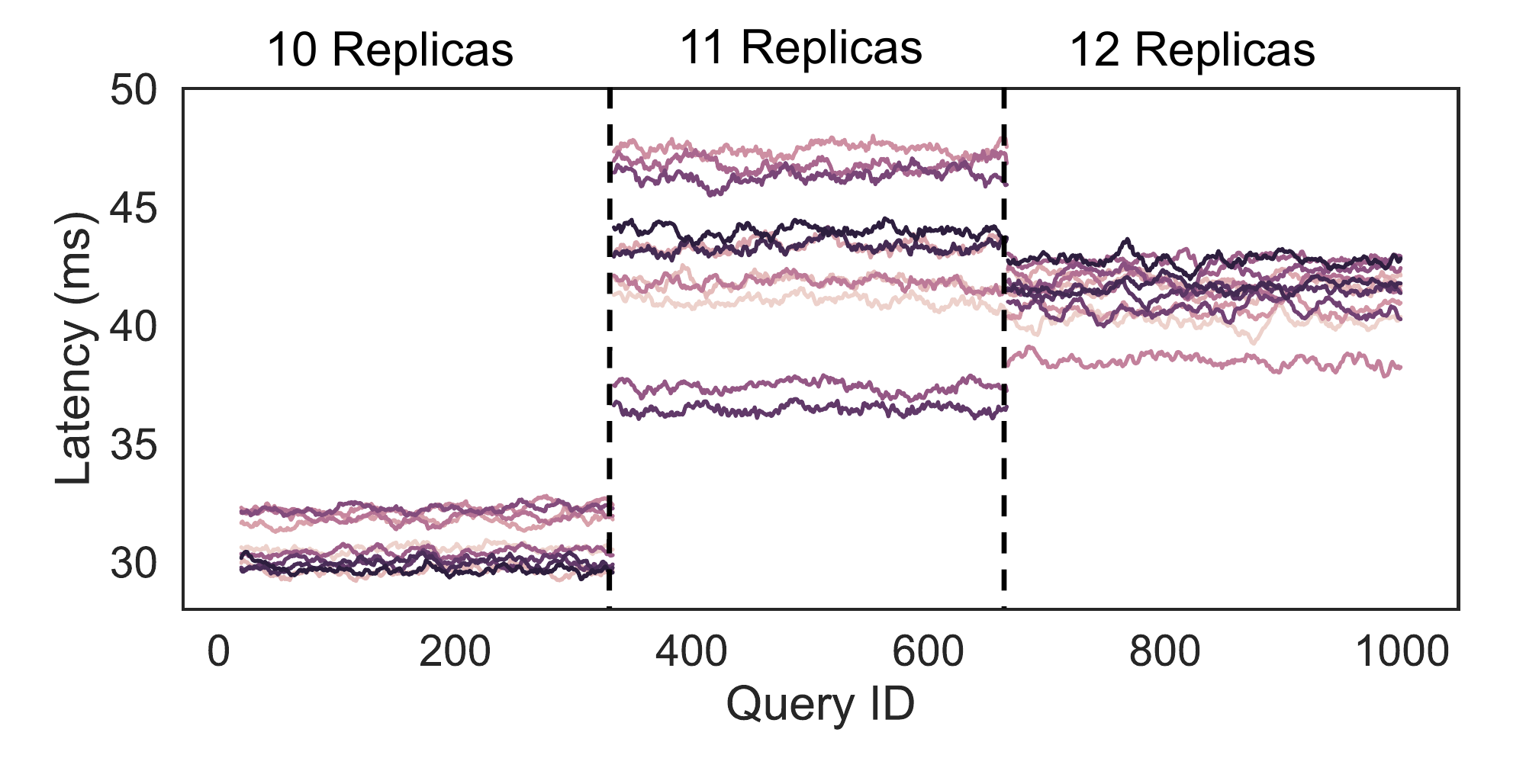}
    \caption{\small{
    Spatial multiplexing on GPUs has unpredictable latency when different number of processes are running concurrently. 
    As we add replicas to a GPU running 10 multi-tenant models, some tenants encounter unpredictable SLO misses.
    A software-level JIT compiler will be able to recognize SLO misses and reprioritize kernels and pick the right execution streams. 
    }}
    \label{fig:eval:unfair}
  \end{figure}





From operating systems~\cite{ritchie1978unix, govil1999cellular} to cluster management~\cite{spark, MapReduce, mesos}, multiplexing is a well-established approach to improve resource utilization.
Systems are usually either time-multiplexed (e.g. CPU core) or space-multiplexed (virtual memory).
Current approaches to multiplex GPUs also fall under these two categories. 
In this section, we discuss why space-only and time-only multiplexing approaches fail to deliver both reliable performance with improved utilization.


\subsection{Inefficient GPU Time Multiplexing}
Each process that interacts with NVIDIA GPU owns a CUDA context. GPU can multiplex multiple CUDA contexts dynamically using an on-device scheduler. 
This approach enables interleaved (but not parallel) execution kernels. Kernels are serialized and processes are preempted periodically. Therefore temporal multiplexing doesn't increase parallelism nor GPU utilization.

As shown in ~\autoref{fig:eval:inefficient}, the inference latency increased linearly as we increase the number of concurrent processes. 
In addition, we noticed the context switching overhead is high because GPUs need to flush the execution pipeline. 


\subsection{Unpredictable Spatial Multiplexing}
Modern GPUs from NVIDIA and AMD can be spatially multiplexed which enables concurrent overlapping kernel execution if resource permits. NVIDIA spatial multiplexing support is provided by Hyper-Q~\cite{nvidia_hyperq} while AMD's MxGPU~\cite{amdmxgpu} utilizes an SR-IOV approach. CUDA Streams and NVIDIA Multi Process Service (MPS)~\cite{nvidiamps} provided application support for spatial multiplexing.
Model inference platforms like ModelBatch~\cite{narayanan2018modelbatch} and NVIDIA TensorRT~\cite{nvidia_tensorrt} utilize CUDA Streams to achieve spatial multiplexing. 

However, these approaches to spatial multiplexing result in poor performance isolation and unpredictable execution times. 
The spatial multiplexing approach is extremely sensitive to the choice of the number of tenants. When there are odd number of tenants on the same GPU, there is greater variability in latency among different processes (\autoref{fig:eval:unfair}).

Finally, because many kernels are tuned assuming they are single-tenant and own the entire GPU, the performance of concurrent execution of such kernels leads to lower throughput (\autoref{fig:solution:retune}).








\section{Proposed solution}
\label{sec:proposal}

We propose dynamic, just-in-time coalescing of execution kernels for GPU inference across multiple streams of execution and over time. This proposal draws inspiration from VLIW compilers by dynamically packing heterogeneous execution kernels to better utilize available hardware resources (spatial dimension).
We also reorder and queue execution kernels in a latency SLO-aware manner to maximize the efficiency of packing. Thus, we advocate for a late-binding, context-aware approach to kernel scheduling on the GPU, in contrast to the early-binding, context-free abstractions currently exposed to the GPU programmer.

We show that the throughput-optimal convolutional block size depends on the size and shape of concurrent blocks of execution; ahead-of-time autotuning informs JIT decisions.
Further, we demonstrate that inter-kernel optimization via coalescing  yields substantial throughput gains.
In the time dimension, reordering queued kernels of execution (a) prioritizes streams with tighter latency budgets, (b) purposefully delays/staggers ill-fitting kernels for better coalescing at a (slightly) later time.


    

    


\subsection{Declarative Kernel Dispatch}
The CUDA programming API is a form of \textit{early-binding}. 
A programmer is required to specify the dimensionality and shape of the program without any contextual information
on the available GPU resources or other kernels of execution that may be in flight. This is inherent to the current low-level set of programming abstractions aimed at proactively controlling how much GPU compute and memory will be utilized. 
We refer to this approach as early-binding and context-free.

In contrast, we believe in \textit{late-binding} and \textit{context-aware} dynamic resource allocation, leveraging runtime information about the number of concurrent kernels of execution, and device context, such as the typical problem sizes served by a device. Given this information, individual kernels can be (a) retuned for better spatial multiplexing, (b) coalesced for better utilization, and (c) reordered for better spatiotemporal packing, while meeting the latency SLOs of individual streams of execution.

Therefore, instead of specifying how GPU should allocate threads across blocks, the programmer should interact with the GPU at a higher level, via a declarative API by specifying the operators, the inputs, and latency constraints. 
The JIT compiler will then execute them with contextual knowledge of the current GPU state and other streams of execution, thereby improving utilization while satisfying latency SLOs.

\begin{figure}[t]
    \centering
    \includegraphics[width=\columnwidth]{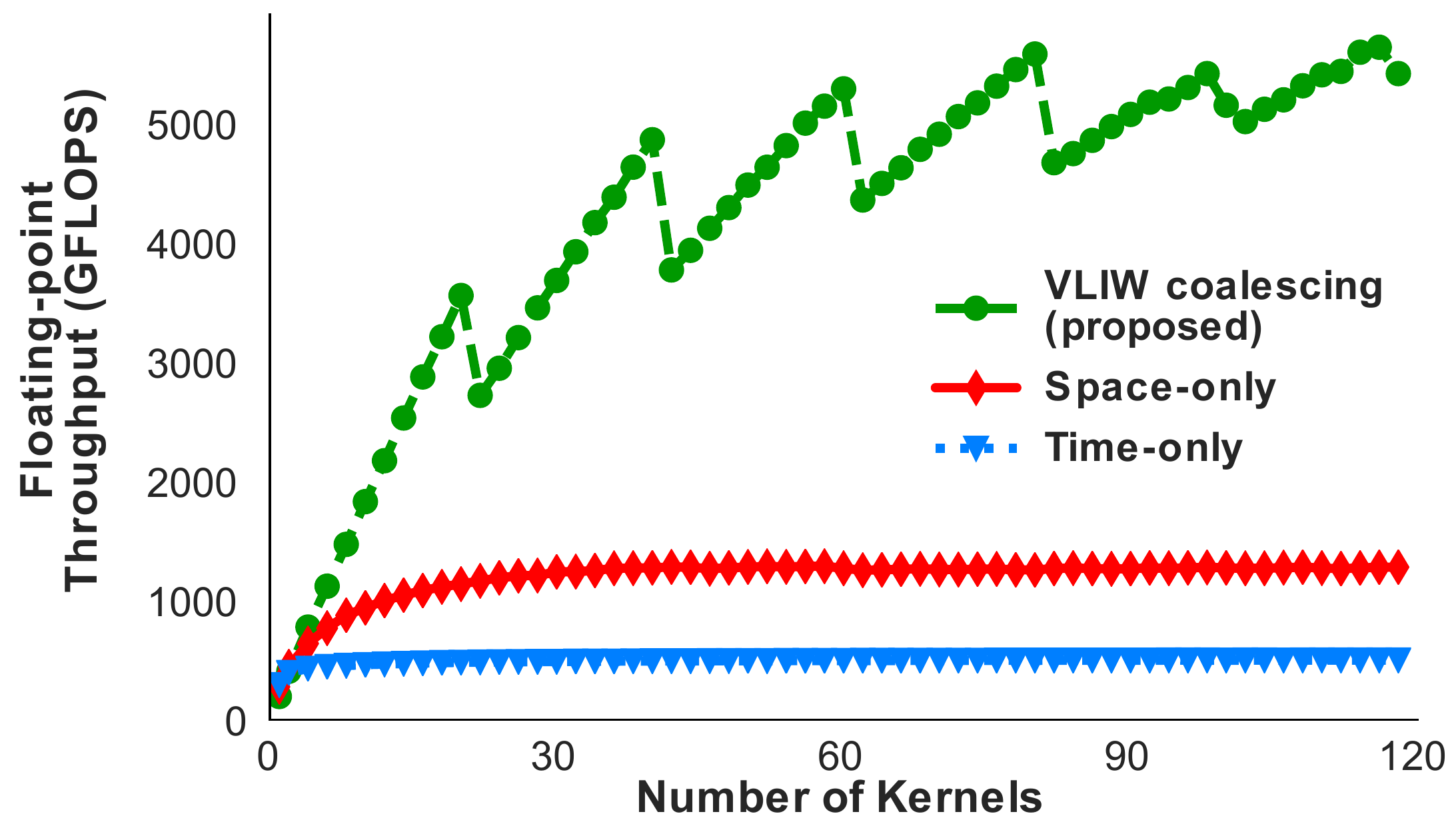}
    \caption{\small{
    Coalesced kernels achieve ideal FP throughput. Coalescing the SGEMM that backs \texttt{conv2\_2} from ResNet-18 with similar problems yields 3.23$\times$ throughput speedup over space-only multiplexing and 7.71$\times$ throughput over time-only multiplexing.
    }}
    \label{fig:res_throughout}
\end{figure}

\subsection{OoO Execution for SLO Attainment}
We preserve predictability and isolation during virtualization by monitoring inference latencies per-kernel. This allows reallocating resources between tenants on-the-fly. Our approach dynamically adjusts to running workloads on the GPU unlike current static compilers like TensorRT~\cite{nvidia_tensorrt} and others~\cite{tensor-comprehension,tvm,halide,GLOW}.

Moreover, we notice that CUDA Stream scheduling anomalies typically only create a few stragglers, so we can simply evict degraded workers without significantly impacting total system throughput. We are further investigating this approach in ongoing work.

A kernel that cannot yet be coalesced with those of other applications can be delayed via reordering. Thus, a dynamic JIT compiler overlaps waiting time from coalescing with computation from other execution streams.

\subsection{VLIW Compilation for Efficiency}
\label{sec:vliw}

\begin{figure}[t]
    \centering
    \includegraphics[width=\columnwidth]{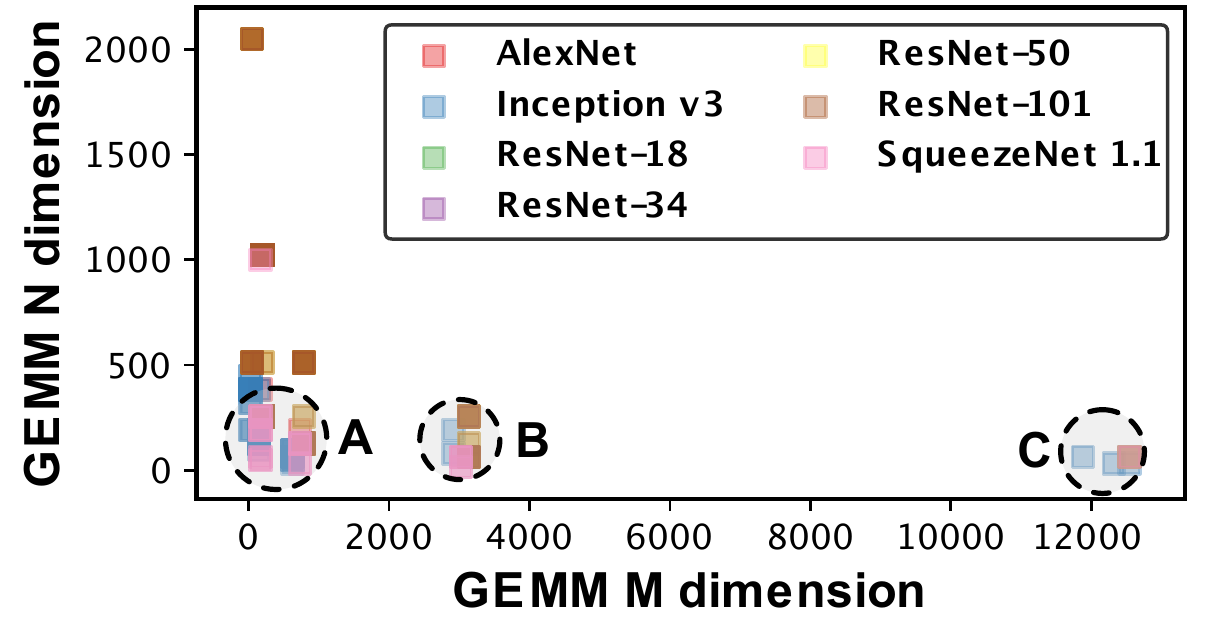}
    \caption{\small{
    Matrix-multiply kernels from a wide class of models are concentrated into several clusters. Problems within certain clusters coalese into efficient super-kernels (A, B and C).
    }}
    \label{fig:gemm_sizes}
\end{figure}

%



Conventional VLIW compilers only modify programs ahead-of-time. However, our dynamic approach uses both ahead-of-time tuning and runtime packing.

By retuning superkernels ahead-of-time in order to improve co-tenancy, we relieve pressure on the runtime JIT compiler. As GPU programs have many tunable parameters, we rely on auto-tuning. Preliminary results for our VLIW auto-tuner show \textit{1.25x throughput gains} in a co-tenancy optimized kernel~(\autoref{fig:solution:retune}).

At runtime, our VLIW JIT compiler repacks multiple small kernels into a large execution block. The compiler can apply pre-computed parameters from the auto-tuning phase to further optimize these larger kernel configurations. Scheduling a single superkernel on the GPU better utilizes compute and memory as compared to time-slicing. Unlike CISC execution streams, DNN kernels have extremely predictable performance and perform a small class of operations; comparably coarse-grained runtime packing decisions are efficient. 

In Figure~\ref{fig:gemm_sizes}, we show that the matrix multiply kernels from multiple frequently used DNNs can be clustered by their dimensions. Within each cluster, problems can be coalesced with minimal padding overhead resulting in efficient co-execution.
Coalescing a ResNet-18 intermediate convolution using the \texttt{cublasSgemmBatched} kernel achieves a geometric mean \textit{7.71$\times$ throughput increase} over time-slicing, and \textit{3.23$\times$} over Hyper-Q spatial multiplexing (Fig.~\ref{fig:res_throughout}). Further, coalescing matrix-vector multiplications common in RNN/LSTM inference yields a 2.48$\times$ throughput speedup over time-slicing \cite{space-time-neurips}. 
VLIW compilation captures this large opportunity gap.

 \begin{table}[]
\begin{tabular}{@{}lll@{}}
\toprule
\multicolumn{1}{c}{\textbf{\begin{tabular}[c]{@{}c@{}}Auto-tuning\\ configuration\end{tabular}}} & \multicolumn{1}{c}{\textbf{\begin{tabular}[c]{@{}c@{}}Uniplexed\\ throughput\end{tabular}}} & \textbf{\begin{tabular}[c]{@{}l@{}}Multiplexed\\ throughput\end{tabular}} \\ \midrule
Greedy kernel & \textit{2.2 TFLOPS} & 4.5 TFLOPS \\
Collaborative kernel & 1.5 TFLOPS & \textit{\textbf{6.1 TFLOPS}} \\ \bottomrule
\end{tabular}
\vspace{2mm}
 

    \caption{
        \small{
            Auto-tuning the blocking configuration on GPU leads to a different type of kernel. 
            \textit{Multi-tenant kernels} achieve 1.25x maximum throughput speedups when dispatched concurrently, despite small (20\%) degradation when run in isolation.
    }
     }
    \label{fig:solution:retune}
    \vspace{-0.2in}
\end{table}

\section{Future Research directions}
\label{sec:future}
With the end of Moore's Law and Dennard Scaling, performance gains in hardware will come from workload specialization. 
This implies increased heterogeneity in the types of specialized devices. 
We envision JIT compilation across multiple streams of execution to extend over multiple devices, such as ASICs, GPUs, TPUs, specialized inference processing units, and FPGAs.
This will enable a more dynamic tradeoff of latency and throughput and improve hardware utilization in heterogenous settings.
Furthermore, dynamic JIT compilation will also be able to explore different latency-accuracy tradeoffs.


\section{Conclusions}
\label{sec:conclude}
The need to execute computationally intensive models within tight interactive latency deadlines has moved DNN inference workloads onto GPUs.
However, the variability in kernel sizes, bursty arrival processes, and tight latency requirements often lead to poor GPU utilization. 
We propose a VLIW-inspired JIT compiler for GPU inference capable of coalescing, reordering, and retuning execution kernels across multiple streams to maximize throughput while meeting latency SLO constraints. 



\section{Acknowledgments}
\label{sec:acks}
We thank Hari Subbaraj and Rehan Sohail Durrani who helped profile kernels as well as Steven Hand, Koushik Sen, Eyal Sela, Zongheng Yang, Anjali Shankar and Daniel Crankshaw for their insightful feedback. In addition to NSF CISE Expeditions Award CCF-1730628, this research is supported by gifts from Alibaba, Amazon Web Services, Ant Financial, Arm, CapitalOne, Ericsson, Facebook, Google, Huawei, Intel, Microsoft, Scotiabank, Splunk and VMware.

\bibliographystyle{ACM-Reference-Format}
\bibliography{fijit}

%
\end{document}